\documentclass[11pt]{article}
\usepackage{epsfig}

\renewcommand{\thefootnote}{\fnsymbol{footnote}}

\begin{document}
\title{\Large \bf 
Off-shell dimensional reduction of 5D orbifold supergravity\footnote{
Talk based on Ref.~\cite{Abe:2007zv} 
given at Summer Institute 2007, Fuji-Yoshida, Japan, August 3-10, 2007}
}

\author{
Hiroyuki~Abe\footnote{E-mail address: 
abe@yukawa.kyoto-u.ac.jp} \\*[10pt] 
{\normalsize \it Yukawa Institute for Theoretical Physics,
Kyoto University, Kyoto 606-8502, Japan }
}

\date{
\centerline{\small \bf Abstract}
\begin{minipage}{0.95\linewidth}
\medskip
\small
We present a systematic way for deriving a four-dimensional 
(4D) effective action of the five-dimensional (5D) orbifold 
supergravity respecting the $N=1$ {\it off-shell} structure. As 
an illustrating example, we derive a 4D effective theory of the 
5D gauged supergravity with a universal hypermultiplet and 
{\it generic} gaugings, which includes the 5D heterotic M-theory 
and the supersymmetric Randall-Sundrum model as special limits 
of the gauging parameters. We show the vacuum structure of 
such model, especially the nature of moduli stabilization, 
introducing perturbative superpotential terms at the orbifold 
fixed points. 
\end{minipage}
}

\maketitle 

\renewcommand{\thefootnote}{\arabic{footnote}}
\setcounter{footnote}{0}

\section{Introduction}
The five-dimensional (5D) gauged supergravity provides interesting 
theoretical frameworks for the physics beyond the standard model 
(SM), e.g., a supersymmetric (SUSY) warped background which can 
realize hierarchical structures of the SM scale and couplings 
(SUSY Randall-Sundrum (RS) model~\cite{Randall:1999ee,Altendorfer:2000rr}), 
an effective theory of the strongly coupled heterotic string~\cite{Horava:1995qa} 
(5D heterotic M-theory~\cite{Lukas:1998yy}), and so on. Therefore, 
It would be important to study generic features of the 5D gauged 
supergravity, and then we present a systematic way for deriving a 
four-dimensional (4D) effective action of such supergravity models. 
We apply it to a simple but illustrating example which includes 
both the SUSY RS model and the 5D heterotic M-theory as special 
limits of the gauging parameters. 

\section{5D gauged supergravity with universal hypermultiplet}
We start from the off-shell formulation on an orbifold~\cite{Kugo:2002js}. 
The 5D superconformal multiplets relevant to our study are the Weyl 
multiplet $\mbox{\boldmath $E$}_W$, the vector multiplets 
$\mbox{\boldmath $V$}^I$ and the hypermultiplets 
$\mbox{\boldmath ${\cal H}$}^{\hat{a}}$, where $I=0,1,2,\ldots,n_V$ 
and $\hat{a}=1,2,\ldots,n_C+n_H$. Here $n_C$, $n_H$ are the numbers 
of compensator and physical hypermultiplets, respectively. 
These 5D multiplets are decomposed into $N=1$ superconformal 
multiplets as $\mbox{\boldmath $E$}_W=(E_W,V_E)$, 
$\mbox{\boldmath $V$}^I=(V^I,\Sigma^I)$ and 
$\mbox{\boldmath ${\cal H}$}^{\hat{a}}=(\Phi^{2 \hat{a}-1},\Phi^{2\hat{a}})$, 
where $E_W$ is the $N=1$ Weyl multiplet, $V_E$ is the $N=1$ general 
multiplet whose scalar component is $e_y^{\ 4}$, $V^I$ is the $N=1$ 
vector multiplet, and $\Sigma^I$, $\Phi^{2 \hat{a}-1}$, $\Phi^{2 \hat{a}}$ 
are $N=1$ chiral multiplets. 

Here we consider the gauged supergravity with 
a single universal hypermultiplet spanning the manifold 
$SU(2,1)/SU(2) \times U(1)$. 
The scalar manifold has an $SU(2,1)$ isometry group, 
which is {\it linearly} realized in the off-shell formulation. 
The situation we consider is realized by taking $(n_C,n_H,n_V)=(2,1,0)$. 
Then the bulk action has a $U(2,1)$ symmetry. The most general 
form of the gauging is parameterized by 
$igt_{I=0} = \sum_{i=1}^8 \tilde\alpha_i \,T^i$ acting on 
$(\Phi^1,\Phi^3,\Phi^5)^t$ or $(\Phi^2,\Phi^4,\Phi^6)^t$, 
where $T^i$ ($i=1,\ldots,8$) are $3 \times 3$ matrix-valued 
generators of $SU(2,1)$. 
In the following, for simplicity, we consider the case that 
$\tilde{\alpha}_i$ are parameterized by three parameters 
($\alpha,\beta,\gamma$) as 
$\tilde\alpha_3 = 2 \beta$, 
$\tilde\alpha_6 = \alpha$, 
$\tilde\alpha_8 = \alpha+\beta+\gamma$, 
$\tilde{\alpha}_i = 0$ ($i \neq 3,6,8$), 
with the corresponding generators in our convention 
$$
T^3=
\left( \begin{array}{ccc}
1 & 0 & 0 \\
0 & -1 & 0 \\
0 & 0 & 0 
\end{array} \right), \qquad 
T^6=
\left( \begin{array}{ccc}
0 & 0 & 0 \\
0 & 0 & 1 \\
0 & -1 & 0 
\end{array} \right), \qquad 
T^8=
\left( \begin{array}{ccc}
0 & 0 & 0 \\
0 & 1 & 0 \\
0 & 0 & -1 
\end{array} \right). 
$$

In the limit $\beta,\gamma\to 0$, this model is reduced 
to the 5D effective theory of heterotic M-theory derived 
in Ref.~\cite{Lukas:1998yy} on-shell. On the other hand, 
in the limit $\alpha \to 0$, we obtain SUSY RS model with 
the AdS curvature scale $k$ and a kink mass $m$ for the 
hypermultiplet, which are related to the gauging parameters as  
$$k=\beta-\gamma/3, \qquad m=3(\beta+\gamma)/2.$$

\section{4D effective theory}
In our model the $Z_2$-even multiplets are $\Sigma^0$, $\Phi^2$, 
$\Phi^3$ and $\Phi^5$. They appear in the action through the 
combinations of $\Sigma^0$, $\Phi^2\Phi^3$ and $\Phi^5/\Phi^3$ 
which carry the zero-modes, i.e., the radion multiplet~$T$, 4D chiral 
compensator~$\phi$ and the universal (chiral) multiplet~$H$, respectively. 
The 5D Lagrangian can be expressed in the $N=1$ 
superspace~\cite{Paccetti:2004ri} as 
\begin{eqnarray}
{\cal L}_{5D} &=& 
-3e^{2\sigma} \int d^4\theta \,
\left( -\partial_y V^0+\Sigma^0+\bar\Sigma^0 \right) 
\Big\{ d_a^{\ b} \bar\Phi^b (e^{-2igt_IV^I})^a_{\ c} \Phi^c \Big\}^{2/3} 
\nonumber \\ &&
-e^{3\sigma} \bigg[ 
\int d^2 \theta\, \Phi^a d_a^{\ b} \rho_{bc} 
(\partial_y-2igt_I\Sigma^I)^c_{\ d} \Phi^d 
+\textrm{h.c.} \bigg] \nonumber\\ && 
+\sum_{\vartheta=0,\pi}
\delta(y-\vartheta R)\,e^{3\sigma}
\bigg\{ \int d^2\theta\,\Phi^2\Phi^3 P_\vartheta(Q) 
+\textrm{h.c.} \bigg\} +\cdots, 
\label{eq:lbulk}
\end{eqnarray}
where $e^{2\sigma}$ is the warp factor of the background geometry, 
$d_a^{\ b}={\rm diag}({\bf 1}_{2n_C},-{\bf 1}_{2n_H})$, 
$\rho_{ab}=i \sigma_2 \otimes {\bf 1}_{n_C+n_H}$ and 
$a,b,\ldots=1,2,\ldots,2(n_C+n_H)$. The ellipsis 
denotes terms including boundary induced K\"ahler potential, 
gauge kinetic function and those of bulk vector multiplets, 
which are all irrelevant to the following discussions. 
We introduce polynomial superpotential terms at the fixed points, 
$$P_\vartheta(Q) = \sum_{n \geq 0} w_\vartheta^{(n)}Q^n,$$ 
where $w_\vartheta^{(n)}$ ($n=0,1,2,\ldots$) are constants, 
and study the nature of moduli stabilization in this model. 

Following the off-shell dimensional reduction procedure 
proposed in Refs.~\cite{Correia:2006pj,Abe:2006eg} (which 
is based on the $N=1$ description of 5D supergravity~\cite{
Paccetti:2004ri} and subsequent studies~\cite{Abe:2005ac}) 
we neglect the kinetic terms for $Z_2$-odd $N=1$ multiplets 
and integrate them out by which the zero-modes of $Z_2$-even 
$N=1$ multiplets are extracted. Then, after integrating the 
bulk Lagrangian over the extra dimension, we obtain the 4D 
effective Lagrangian 
\begin{eqnarray}
{\cal L}_{4D} &=& -3 \int d^4\theta \, 
|\phi|^2 e^{-K/3} 
+\left\{ \int d^2 \theta \, \phi^3 W 
+\textrm{h.c.} \right\}, 
\nonumber
\end{eqnarray}
with the K\"ahler potential~$K$ and the superpotential~$W$ 
given by~\cite{Abe:2007zv} 
\begin{eqnarray}
K &=& -3 \ln \int^{\pi {\rm Re}\,T}_0 dt\,
e^{-2 \beta t} \bigg\{ 
\cosh(2t \tilde\gamma)\,(1-|H|^2) 
\nonumber \\ && \qquad 
+\sinh(2t \tilde\gamma)\,
\frac{(\alpha+\gamma)(1+|H|^2)+\alpha 
(H+\bar{H})}{\tilde\gamma}
\bigg\}^{\frac{1}{3}}, 
\nonumber \\
W &=& \frac{1}{4}\sum_{\vartheta=0,\pi} e^{-3\vartheta \beta T} 
\left\{ c_\vartheta+(\alpha+\gamma) s_\vartheta 
+\alpha s_\vartheta H \right\} 
P_\vartheta(H_\vartheta), 
\nonumber
\end{eqnarray}
where 
$$
H_\vartheta=
\frac{-\alpha s_\vartheta
+\left( c_\vartheta 
-(\alpha+\gamma) s_\vartheta \right) H}{
c_\vartheta +(\alpha+\gamma) s_\vartheta 
+\alpha s_\vartheta H}, 
$$
and 
$c_\vartheta= 
\cosh(\vartheta T \tilde\gamma)$, 
$s_\vartheta= 
\sinh(\vartheta T \tilde\gamma)/\tilde\gamma$, 
$\tilde\gamma=\sqrt{\gamma^2 -2\alpha \gamma}$. 

\begin{figure}[t]
\begin{center}
\ \\*[-30pt]
\begin{minipage}{0.48\linewidth}
\begin{center}
\epsfig{file=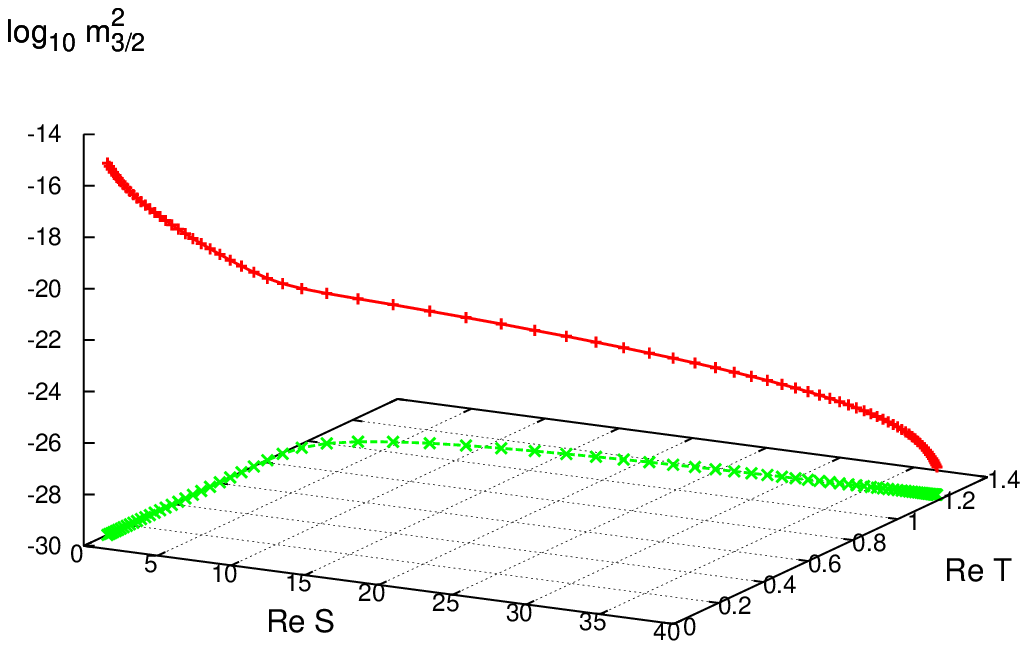,width=1.1\linewidth} 
\\*[-20pt] (a) 
\end{center}
\end{minipage}
\hfill
\begin{minipage}{0.48\linewidth}
\begin{center}
\epsfig{file=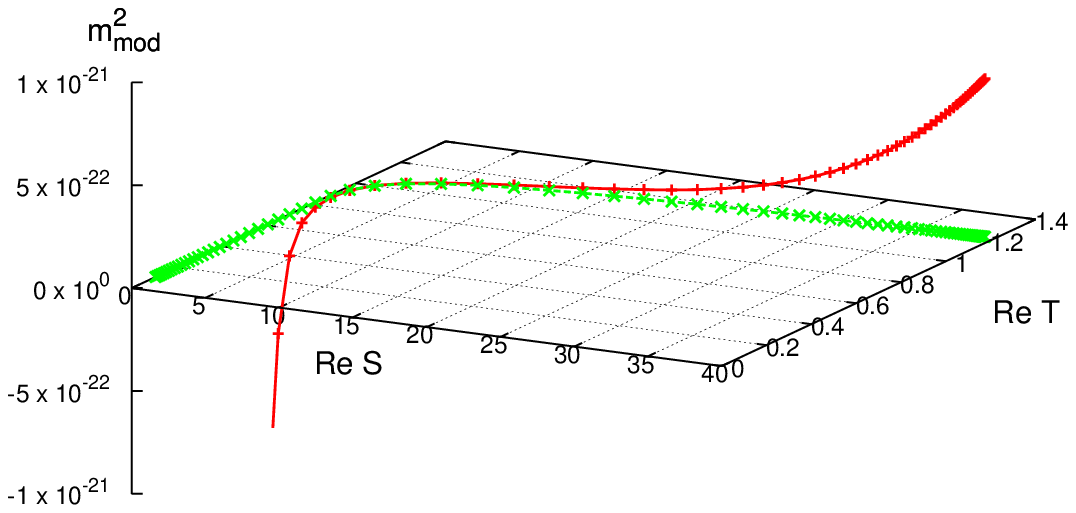,width=1.1\linewidth} 
\\[-20pt] (b) 
\end{center}
\end{minipage} \\*[-20pt] \ 
\end{center}
\caption{
(a) The gravitino and (b) the lightest moduli mass square 
at the SUSY point on the $({\rm Re}\,S,{\rm Re}\,T)$-plane. 
Here $S=(1-H)/(1+H)$ and we restrict the parameters as 
$w_\vartheta^{(0)}/w_\vartheta^{(1)}=(c+1)/(c-1)$ 
for $\vartheta=0,\pi$. 
We vary the parameter $\alpha$ from $100$ (left end-point) to 
$0.1$ (right end-point) by fixing 
$\beta=1$, 
$c=-40$, 
$w_\pi^{(0)}/w_0^{(0)}=10^5$ 
and $w_0^{(0)}-w_0^{(1)}=8 \times 10^{-14}$. 
All the mass scales are measured in the unit $M_{\rm Pl}=1$.
(From Ref.~\cite{Abe:2007zv}.)}
\label{fig:stm}
\end{figure}

In the limit $\gamma \to 0$, the $t$-integration in 
the above K\"ahler potential is carried 
out analytically. In this case, we find a SUSY 
preserving extremum of the scalar potential with 
constant and tadpole superpotential terms at the 
fixed points ($w_\vartheta^{(n \ge 2)}=0$). 
The gravitino and the lightest modulus mass squares, 
$m_{3/2}^2$ and $m_{\rm mod}^2$, at this SUSY point 
with various values of $\alpha$ is shown in 
Fig.~\ref{fig:stm} (a) and (b), respectively. 
We find that $m_{3/2}^2$ ($m_{\rm mod}^2$) 
monotonically decreases (increases) as $\alpha$ 
decreases. 

On the other hand, in the limit $\alpha \to 0$, our model 
is reduced to SUSY RS model with the AdS curvature $k$ 
and an independent kink mass $m$ shown before. 
With the generic superpotential $w_\vartheta^{(n)} \ne 0$, 
the hypermultiplet is stabilized at the origin $H=0$, 
where the effective theory of the radius modulus $T$ 
is described by 
$$
K=-3 \ln \frac{1-e^{\pi k(T+\bar{T})}}{2k}, \qquad 
W=\frac{1}{4}(w_0^{(0)}+w_\pi^{(0)}e^{-3 \pi kT}). 
$$
If the parameters satisfy 
$-w_0^{(1)}/w_\pi^{(1)}
=(-w_0^{(0)}/w_\pi^{(0)})^{m/k+3/2}$, 
we easily find a SUSY AdS minimum at 
$\pi kT=\ln (-w_\pi^{(0)}/w_0^{(0)})$ 
where the gravitino and moduli mass square are found as 
$m_{3/2}^2=k^3(w_0^{(0)})^2 
\left\{ 1-(w_0^{(0)})^2/(w_\pi^{(0)})^2 \right\}^2/2$ and 
$(m^2_{T+},m^2_{T-})=(4m_{3/2}^2,0)$, respectively.

\section{Summary}
Based on the off-shell dimensional reduction~\cite{Correia:2006pj,Abe:2006eg}, 
we studied a 4D effective theory of the 5D gauged supergravity 
with a universal hypermultiplet and {\it generic} gaugings and 
found SUSY preserving extrema with perturbative superpotential 
terms at the orbifold fixed points. These results would be 
useful for further studies of moduli stabilization and the 
uplifting of AdS minima in this class of models~\cite{Abe:2007zv,Correia:2006vf}.

\subsection*{Acknowledgement}
The author would like to thank Yutaka~Sakamura 
for the collaboration~\cite{Abe:2007zv} 
which forms the basis of this talk, 
and also the organizers of Summer Institute 2007. 
The author is supported by the Japan Society for the Promotion 
of Science for Young Scientists (No.182496).

\end{document}